\documentclass[conference]{IEEEtran}
\IEEEoverridecommandlockouts
\usepackage{cite}
\usepackage{amsmath,amssymb,amsfonts}
\usepackage{algorithmic}
\usepackage{graphicx}
\usepackage{textcomp}
\usepackage{xcolor}
\usepackage[hidelinks]{hyperref}

\def\BibTeX{{\rm B\kern-.05em{\sc i\kern-.025em b}\kern-.08em
    T\kern-.1667em\lower.7ex\hbox{E}\kern-.125emX}}
\begin{document}

\title{Active Learning Pipeline for Brain Mapping in a High Performance Computing Environment\\
\thanks{DISTRIBUTION STATEMENT A. Approved for public release. Distribution is unlimited. This material is based upon work supported by the Defense Advanced Research Projects Agency under Air Force Contract No. FA8702-15-D-0001. Any opinions, findings, conclusions or recommendations expressed in this material are those of the author(s) and do not necessarily reflect the views of the Defense Advanced Research Projects Agency. This work was also sponsored by the National Institutes of Health NIH 1U01MH117072. Any opinions, findings, conclusions or recommendations expressed in this material are those of the author(s) and do not necessarily reflect the views of the National Institutes of Health. The work performed by University of Florida was sponsored by the Defense Advanced Research Projects Agency (DARPA) BTO under the auspices of Dr. Douglas Weber and Dr. Tristan McClure-Begley through the DARPA Contracts Management Office Grant No. HR0011-17-2-0019.}
}

% This is the author block that will be displayed
\author{
\IEEEauthorblockN{
Adam Michaleas\IEEEauthorrefmark{1}, 
Lars A. Gjesteby\IEEEauthorrefmark{1}, 
Michael Snyder\IEEEauthorrefmark{1}, 
David Chavez\IEEEauthorrefmark{1}, 
Meagan Ash\IEEEauthorrefmark{2},\\ 
Matthew A. Melton\IEEEauthorrefmark{2}, 
Damon G. Lamb\IEEEauthorrefmark{2}, 
Sara N. Burke\IEEEauthorrefmark{2}, 
Kevin J. Otto\IEEEauthorrefmark{2},\\
Lee Kamentsky\IEEEauthorrefmark{3}, 
Webster Guan\IEEEauthorrefmark{3}, 
Kwanghun Chung\IEEEauthorrefmark{3}, 
Laura J. Brattain\IEEEauthorrefmark{1}}

\IEEEauthorblockA{\IEEEauthorrefmark{1}MIT Lincoln Laboratory, Lexington, MA, USA\\
\{adam.michaleas, lars.gjesteby, michael.snyder, david.chavez, brattainl\}@ll.mit.edu}
\IEEEauthorblockA{\IEEEauthorrefmark{2}University of Florida, Gainesville, FL, USA\\
\{meaganash, mmelton2, dlamb, burkes, kevin.otto\}@ufl.edu}
\IEEEauthorblockA{\IEEEauthorrefmark{3}Massachusetts Institute of Technology, Cambridge, MA, USA\\
\{lkaments, wjguan, khchung\}@mit.edu}
}

\maketitle

\begin{abstract}
This paper describes a scalable active learning pipeline prototype for large-scale brain mapping that leverages high performance computing power. It enables high-throughput evaluation of algorithm results, which, after human review, are used for iterative machine learning model training. Image processing and machine learning are performed in a batch layer. Benchmark testing of image processing using pMATLAB shows that a 100$\times$ increase in throughput (10,000\%) can be achieved while total processing time only increases by 9\% on Xeon-G6 CPUs and by 22\% on Xeon-E5 CPUs, indicating robust scalability. The images and algorithm results are provided through a serving layer to a browser-based user interface for interactive review. This pipeline has the potential to greatly reduce the manual annotation burden and improve the overall performance of machine learning-based brain mapping.  
\end{abstract}


\section{Introduction}
One of the top priorities of the BRAIN Initiative led by the US Government is to map human brains at multiple scales (\href{https://braininitiative.nih.gov/}{https://braininitiative.nih.gov/}). Detailed maps of connected neurons in both local circuits and distributed brain systems, once reconstructed, will facilitate our understanding of the relationship between neuronal structure and function. Advances in brain imaging techniques have made it possible to image the brain structures at high throughput (on the order of terabytes/hour), over a large field of view (multiple brain regions), and at high resolution (cellular or sub-cellular)~\cite{silvestri_confocal_2012,chung2013clarity,osten_mapping_2013,tomer_advanced_2014,keller_visualizing_2015,stefaniuk_light-sheet_2016}. Datasets of a whole human brain are estimated to be on the order of up to several petabytes, which is effectively impossible to process manually. Image processing and visualization techniques are being developed to assist the neuroscientific discovery~\cite{fornito_fundamentals_2016,peng_extensible_2014,peng_bigneuron_2015}. While there are many methods to analyze high-resolution neuroimaging data, accurate neuron segmentation and tracing at scale (terabyte and above in size) are some of the fundamental processing tasks that still need to be optimized. 

\section{Related Work}
\subsection{Existing deep learning algorithms for neuron segmentation}
Deep learning-based methods, such as convolutional neural networks, often use segmentation techniques to distinguish neurons from the background \cite{li20193d,li2017deep,wang2019multiscale}. Two architectures in particular, Mask Regional CNN (R-CNN)~\cite{he2017mask, johnson2018adapting} and 2D or 3D U-net~\cite{ronneberger2015u, cciccek20163d}, are gaining popularity in this area. Vuola \textit{et al.}\cite{vuola2019mask} examined the strengths and weaknesses of the two techniques on the nuclei segmentation data from the Kaggle 2018 Data Science bowl\cite{ljosa2012annotated}. Various improvements have been suggested for the use of 3D U-Net to trace axons\cite{wang2019multiscale}. Flood-filling to trace neurons outward from an initial neuron voxel~\cite{januszewski2018high} has also been reported. Our recent work has explored transfer learning and domain adaptation methods for neuron segmentation with limited annotations~\cite{khorrami2019deep}. 

\subsection{Active learning for brain mapping}

\begin{figure*}[htbp]
  \includegraphics[width=\textwidth,height=8cm]{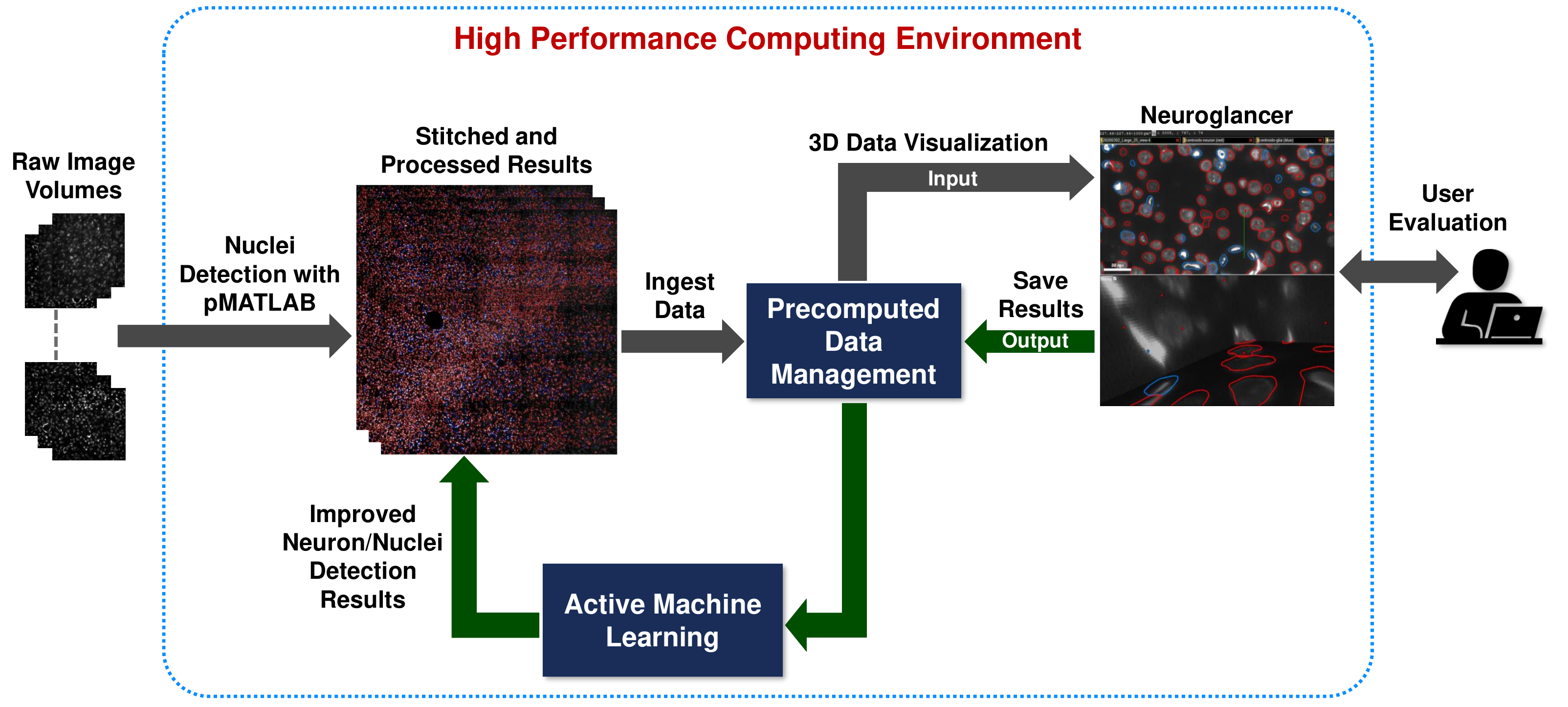}
  \caption{Overview of the proposed active learning pipeline in an HPC environment.}
  \label{fig_pipeline}
\end{figure*}

While deep learning-based approaches have shown to be effective in neuron segmentation and tracing, one major challenge is the lack of annotated data, which often requires domain knowledge. The manual process is also laborious and time consuming. In addition, there is lack of tools that allow domain experts to review the algorithm results at scale. 

In the scenarios where data may be abundant but labels are scarce or expensive to obtain, active learning is a viable solution and has been used in modern machine learning. Active learning is a special case of machine learning in which a learning algorithm can interactively cue a user to label new data points with the desired outputs. Active learning aims to achieve high accuracy using as few labeled instances as possible, thereby minimizing the cost of obtaining labeled data~\cite{settles_active_2009}. How to select data samples for human annotation is an ongoing research area, where a variety of algorithms have been investigated~\cite{yang2017suggestive,kuo2018cost}. Another key component of an active learning process is the interactive user interface through which a human annotator reviews the selected queries. Even though there are open source neuroscientific software~\cite{schneider2012nih,peng_extensible_2014} and commercial products such as Neurolucida 360 (MBF Bioscience, Williston, VT) available, existing tools are primarily workstation-based and are limited to processing a few neurons at a time. Thus, they are not scalable to densely labelled microscopy brain data of multiple brain regions. 

To address this unmet need, we developed a prototype active learning pipeline in a high performance computing (HPC) environment, which enables parallel data processing tasks (including image processing and machine learning) behind the scenes, while supporting an interactive browser-based front-end graphical user interface with 2D/3D views. Users can review the images and algorithm outputs through the browser interface. Modification of the algorithm outputs are saved and sent back to the data management server for the next iteration of machine learning model training. Fig.~\ref{fig_pipeline} illustrates the main components of the pipeline. 

\section{Materials and Methods}

\subsection{Dataset}
In this paper, we will primarily focus on data acquired using microscopy (e.g., light-sheet) from brain tissue samples prepared with tissue clearance (e.g., CLARITY~\cite{chung2013clarity}) and labeled with dense-labeling markers. Each dataset, composed of multiple partially overlapping z-stacks that can be registered into a consolidated 3D volume, often reaches multiple terabytes in size. To develop and test our pipeline, we used a dataset acquired from the cortex region of a rat brain as an example. The tissue sample measured approximately 4mm~$\times$~3mm~$\times$~2mm, and free floating sections went through immunofluorescence staining and labeling of cell nuclei. A fluorescent DNA stain, DAPI (4’,6-diamidino-2-phenylindole), was used to mark nuclei, and a protein stain was used to tag the proto-oncogene c-Fos that is expressed within some neurons following neuronal activity~\cite{morgan1987mapping}. A light-sheet microscope (ZEISS Lightsheet Z.1, Oberkochen, Germany) with a 20$\times$ objective acquired images of the sample in a 5~$\times$~5 snake-like grid pattern, yielding 25 separate image volumes. Each volumetric stack consisted of 1920~$\times$~1920~$\times$~397 voxels (volumetric pixels) at a resolution of 0.227~\textmu m~$\times$~0.227~\textmu m~$\times$~1~\textmu m. The data were stored as 16-bit grayscale slices on two channels corresponding to the signals from each of the two stains. As the microscope sweeps across the sample, there is typically 5\% to 10\% of overlap between adjacent locations in the grid so that cells are not clipped at the edges of the field of view. Thus, a stitching algorithm is needed in post-processing to account for variable overlap when merging the image volumes back together.

\subsection{High performance computing environment}
We used the MIT Supercloud as the HPC environment for our pipeline prototype development. The MIT Supercloud enables traditional enterprise computing and cloud computing workloads to be run on an HPC cluster. The software stack, which contains all of the system and application software, resides on every node~\cite{prout2017supercloud}.

HPC systems require efficient mechanisms for rapidly identifying available computing resources, allocating those resources to programs, and launching the programs on the allocated resources. The open-source SLURM software~\cite{yoo2003slurm} provides scalable cluster management, a job scheduling system, and is independent of programming language or parallel programming models~\cite{prout2017supercloud}.

One unique tool on the MIT Supercloud is parallel MATLAB (pMATLAB), which parallelizes MATLAB scripts by implementing parallel global array semantics using standard operator overloading techniques~\cite{travinin2007pmatlab}. This allows scientists and engineers to quickly prototype algorithms in MATLAB and launch the runs on HPC through SLURM job scheduling without having to acquire in-depth parallel programming knowledge beforehand. 

\subsection{Proposed active learning pipeline}

\iffalse % START block comment
\begin{figure}[t]
    \includegraphics[width=\columnwidth, height=2cm]{NG_layers1.png}
    \caption{Channels of brain imagery are mapped to layers in Neuroglancer. All of the layers are displayed by default. The user can choose to turn off a layer by clicking on the red X button.}
    \label{fig_NG_layers}
\end{figure}
\fi % END block comment

In our proposed active learning pipeline, neuron detection algorithms are applied to brain imaging datasets using pMATLAB so that all the volumes in a dataset are processed at the same time. The raw images and results are then stitched together to form the original large field of view. The stitched data and results are ingested into a data management server called a precomputed server (PCS) (\href{https://github.com/chunglabmit/precomputed-tif}{https://github.com/chunglabmit/precomputed-tif}), initially developed by the Chung Lab at MIT. The data and results can then be served/queried through a browser-based tool called Neuroglancer  (\href{https://github.com/google/neuroglancer}{https://github.com/google/neuroglancer}), originally developed by Google.

We made a number of modifications and extensions to the PCS and Neuroglancer in order to ingest, serve, and visualize raw images and algorithm results, as well as save any changes made by users for iterative machine learning model training:

\begin{itemize}
\item Ingest raw image volumes and neuron/glia/centroid/axon algorithm outputs
\item Support visualization of algorithm detections overlaid on the raw imagery (along with existing visualization of imagery)
\item Save updated annotation data after human review
\item Provide data scalability (for use with Neuroglancer) by breaking up annotation data into blocks and keeping stitched imagery
\item Support JSON, CSV, and HDF5 formats 
\item Support serving multiple datasets at once
\end{itemize}

\begin{figure}[t]
    \includegraphics[width=\columnwidth]{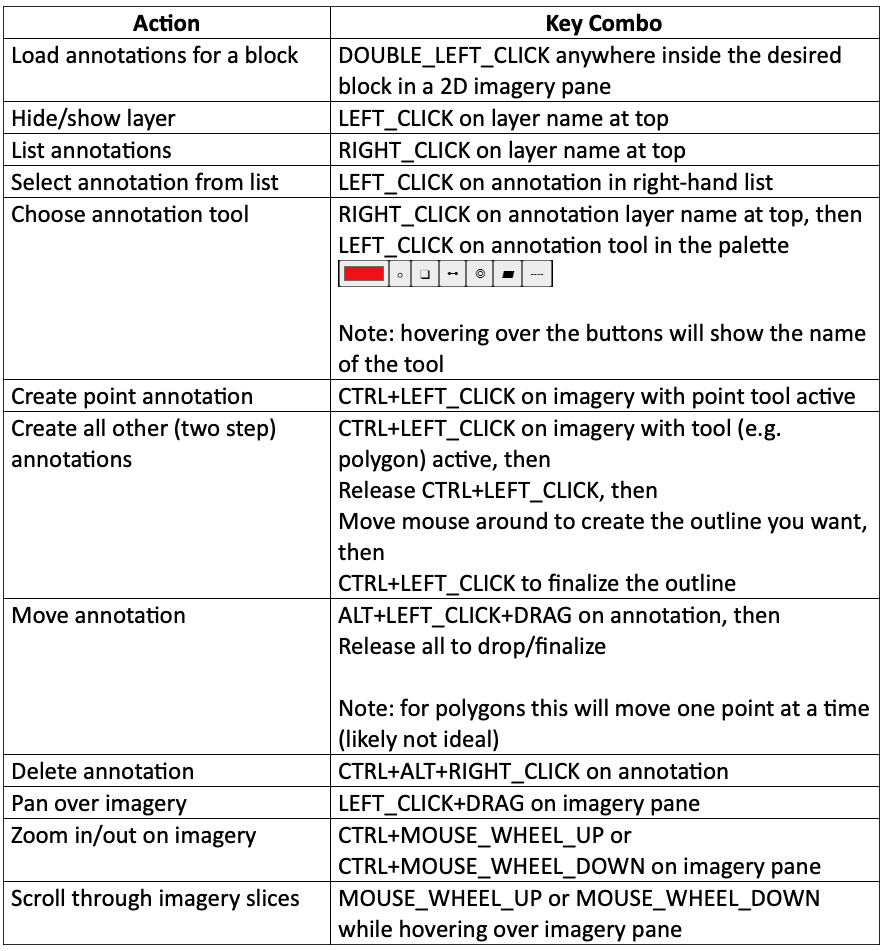}
    \caption{List of major functions extended/implemented in Neuroglancer for data evaluation.}
    \label{fig_NG_functions}
\end{figure}

We augmented the concept of layers in Neuroglancer to display image data from each channel (e.g., images from multiple stains such as DAPI and c-Fos) and algorithm results in layers, which can be turned on and off dynamically by the user. All of the layers are displayed by default. We also extended existing functions or implemented new features in Neuroglancer to facilitate the review of data. Fig.~\ref{fig_NG_functions} is a list of major actions and key combos that we extended/developed for human evaluation.

\subsection{Batch Layer}

\begin{figure*}
    \centering
    \includegraphics[width=0.9\linewidth]{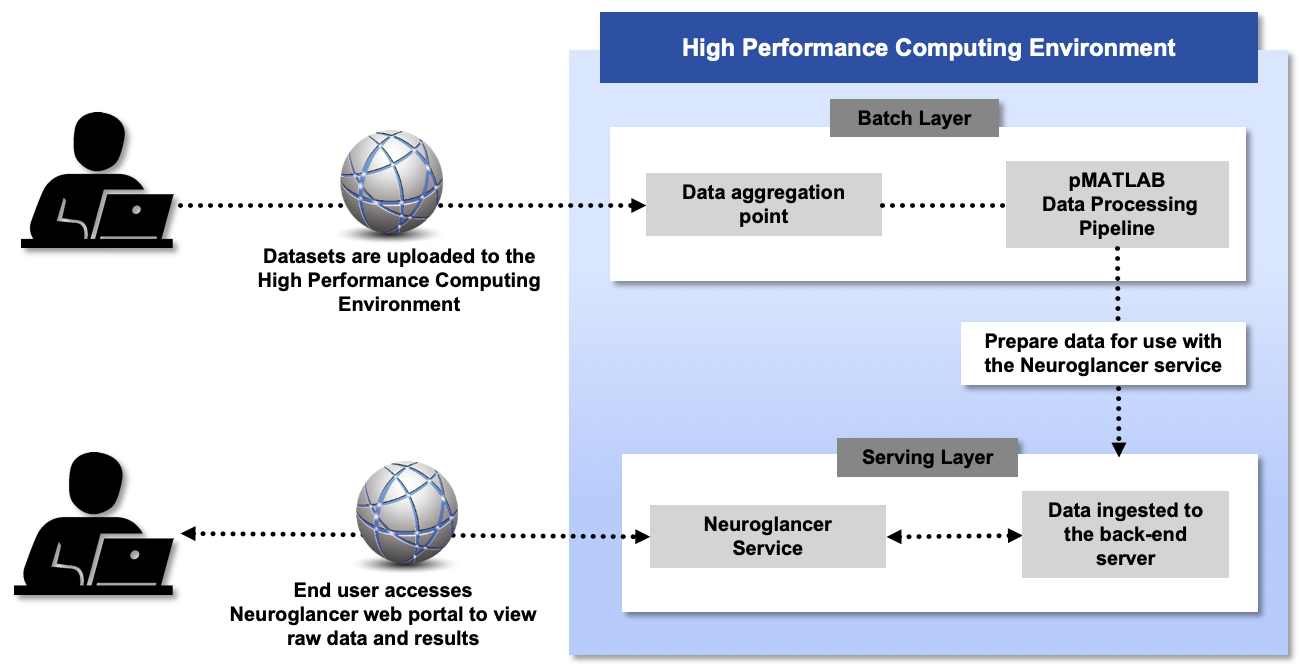}
    \caption{Flow diagram of the batch and serving layers.}
    \label{fig_batch_serving_layers}
\end{figure*}

Fig.~\ref{fig_batch_serving_layers} provides a more detailed view of the HPC environment, which is divided into a batch layer and a serving layer. The batch layer performs the image processing that generates results to be reviewed by the end user. For the example dataset used in this paper, the batch layer first uses pMATLAB code to perform automatic 3D segmentation of cell nuclei in each image volume in the 25-block dataset, which contains approximately $37\times10^9$ voxels. Our segmentation method employs the difference of Gaussians technique in combination with a 3D morphological watershed algorithm to extract nuclei edges and segment individual regions~\cite{meyer1994topographic}. The next step is to separate neurons and glial cells within the cortex images in order to perform multi-channel coincidence analysis on neurons only. The signals from the DAPI stain correspond to nuclei locations of all cell types within the cortex. In addition to neurons, this brain region includes glia, such as astrocytes, oligodendrocytes, and Schwann cells. After 3D segmentation, a support vector machine (SVM) was applied to split the cell nuclei detection into neuron and glia classes. The SVM was previously trained on an image volume containing manually labeled cells with several features, including volume, diameter, and statistical measures of voxel intensity (mean, standard deviation, kurtosis, and skew). Through five-fold cross-validation, the SVM achieved a mean accuracy of 97\% as measured by area under the receiver operating characteristic (ROC) curve. After nuclei detection is completed throughout each image volume, the dataset is stitched back together in the original grid pattern, with overlapping regions automatically detected and merged with our dynamic stitching algorithm. This algorithm searches for the optimal overlap between neighboring image volumes in two dimensions. The images are initially aligned with one voxel of overlap starting at the edges, and then stepped in one voxel increments up to 10\% of their width and height. A difference image of each overlap region is computed to serve as a loss function, which is normalized by the total number of voxels in the region. The minimum loss value occurs at the optimal number of overlapping voxels.

\subsection{Serving Layer}

The serving layer of this pipeline consists of a browser-based front-end interface (Neuroglancer) used for visualization of brain imagery (e.g., axons and neuron centroids) and the PCS services which acts as a back-end for hosting the images for use with Neuroglancer. HAProxy is used to enable the web proxy capabilities for HTTPS requests for the Neuroglancer and PCS services. All of these components of the serving layer are started using a single master SLURM batch script in the HPC environment.

\section{Results}
\begin{figure*}
\includegraphics[width=\textwidth]{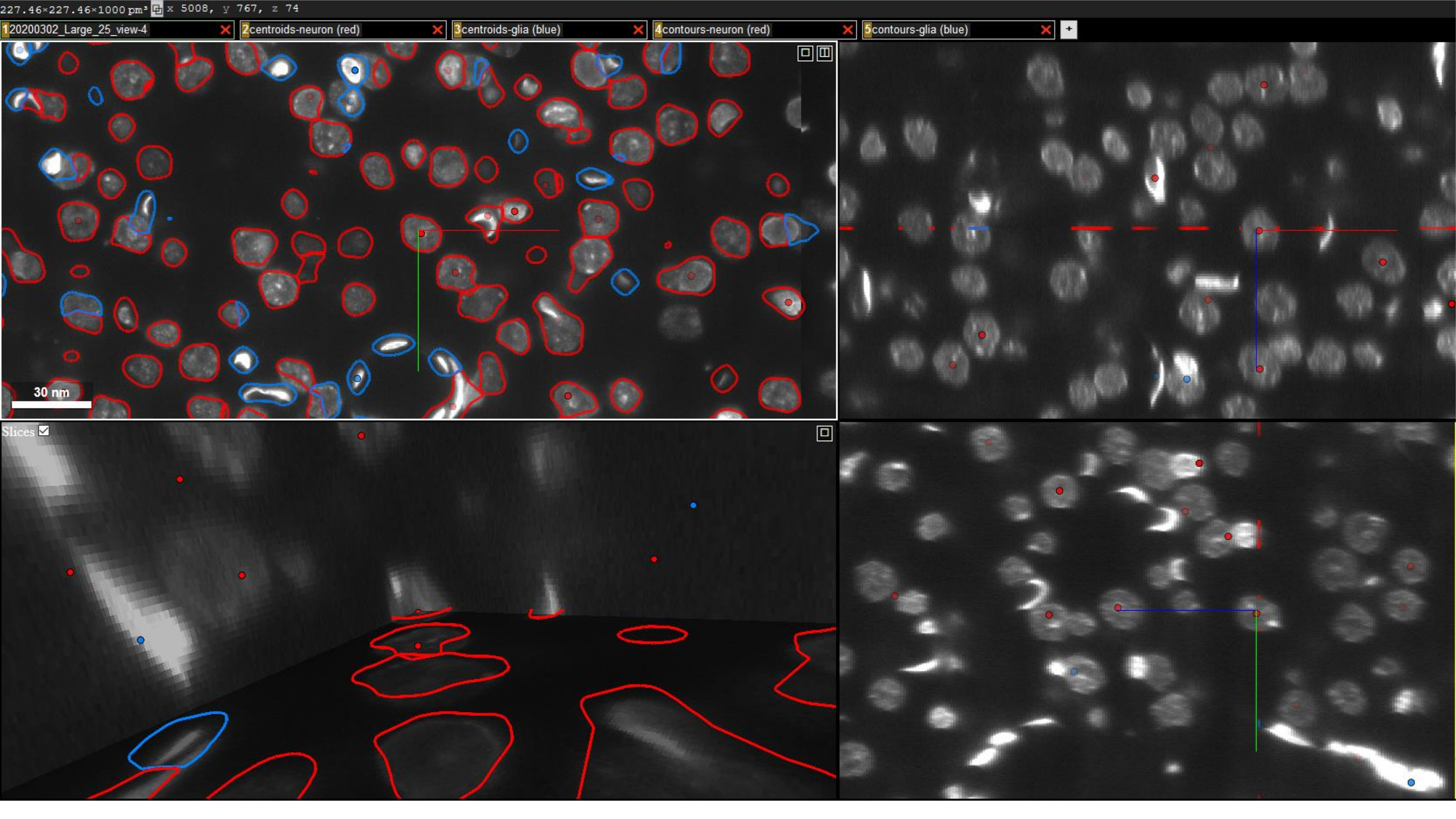}
\caption{Example of the DAPI channel in Neuroglancer. The four-pane view consists of 3 orthogonal cross-sectional views as well as a 3D view (bottom left pane). Raw data are displayed as grayscale images in 3D. The algorithm segmented neurons are outlined in red, and the detected centroids are marked with red dots. Glial cells are marked in blue.}
\label{fig_NG}
\end{figure*}

Fig.~\ref{fig_NG} is an example of the DAPI channel raw imagery and algorithm results displayed in the Neuroglancer interface. The four-pane view consists of 3 orthogonal cross-sectional views as well as a 3D view (bottom left pane). Raw data imagery is displayed as grayscale images in 3D. The algorithm-segmented neurons are outlined in red, and the detected centroids are marked with red dots. Glial cells are marked in blue. 

In this example, there are 25 volumes/blocks of raw images, and all are displayed to provide a global overview. To maintain the interactive nature of the interface, only the algorithm results from the block where the mouse cursor is positioned will be overlaid on top of the raw imagery. The user can zoom and pan, scroll through the slices, and move, add, or delete an annotation. The header at the top of the Neuroglancer interface displays the information of the dataset and annotation layers. All of the layers are displayed by default. The user can choose to turn off a layer by clicking on the name.

\begin{figure}
\includegraphics[width=\columnwidth]{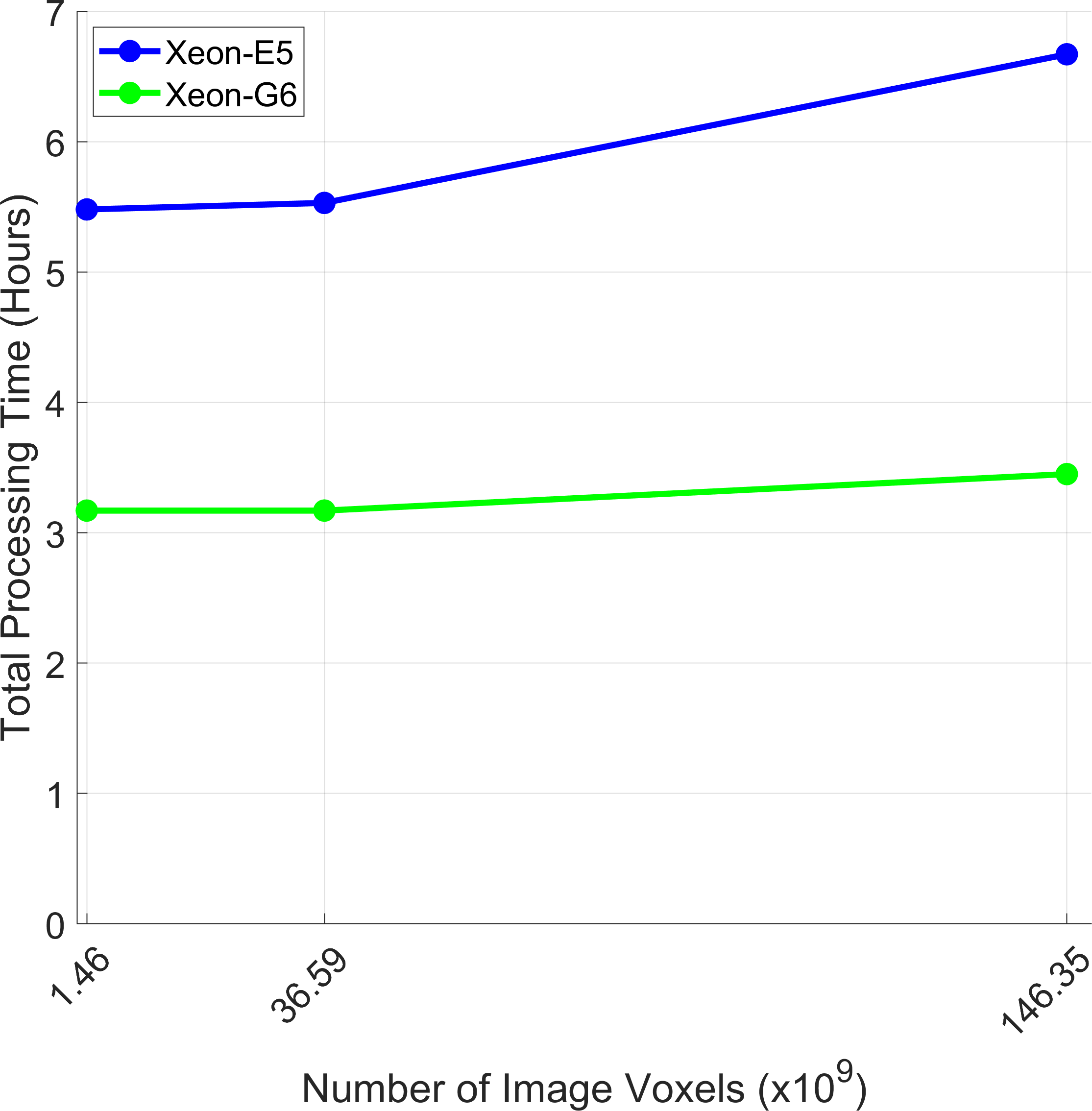}
\caption{Performance evaluation of the batch layer in parallel processing 1, 25, and 100 image volumes with pMATLAB, reported as the number of image voxels. The time difference to process 1 and 100 image volumes is within 0.5 hours on the Xeon-G6 CPUs and about one hour on the Xeon-E5 CPUs, indicating robust scalability.}
\label{fig_performance_plot}
\end{figure}

Fig.~\ref{fig_performance_plot} shows the performance benchmarking results of the timing and scalability of the parallel image processing pipeline in the batch layer. We assessed the scenarios of processing a single image volume, a 5~$\times$~5 grid (25 volumes), and a 10~$\times$~10 grid (100 volumes) using pMATLAB, and reported these sizes as the total number of voxels processed. The experiments were repeated on two CPU types: Intel Xeon CPU E5-2680 v4 (abbreviated as Xeon-E5) and Intel Xeon Gold 6248 CPU (abbreviated as Xeon-G6). The plot shows how utilizing an HPC environment with parallelized code can generate results at large scales without a significant increase in time cost. The trend holds up on both the Xeon-E5 and Xeon-G6 CPUs. Specifically, we achieved a 100$\times$ increase in throughput (10,000\%) while only increasing the total processing time by 9\% on the Xeon-G6 and by 22\% on the Xeon-E5.

\section{Summary and Future Work}
This paper presented a scalable active learning pipeline prototype for large-scale brain mapping that leverages high performance computing power. Image processing and machine learning are performed in a batch layer, while images and algorithm results are provided through a serving layer to a browser-based user interface for interactive review. This pipeline has the potential to greatly reduce the manual annotation burden and improve the overall performance of machine learning-based brain mapping.

With the active learning pipeline prototype established, we will process more large-scale brain imaging datasets in the batch layer, visualize the neuron detection results in the serving layer for user review, and then incorporate the corrected segmentation maps as labels for re-training machine learning algorithms. Moving towards imaging of a whole rat brain, we could expect data acquisition on the order of a 100~$\times$~100 grid, yielding 10,000 image volumes. At our current scaling rate on Xeon-G6 CPUs, assuming resource availability, we estimate that a 10,000$\times$ increase in throughput (1,000,000\%) over a single volume would only cost an additional 30\% in computing time. More functions in Neuroglancer will be implemented to facilitate the human evaluation. We will also explore advanced algorithms, including query strategy frameworks~\cite{settles_active_2009}, to optimize the selection of data samples to be reviewed by the user in order to maximize the learning outcome with as few number of samples to review as possible. Over time, we expect this active learning framework to yield increasingly better accuracy in segmenting and tracing neurons in unlabeled image volumes.

\section*{Acknowledgment}

The authors wish to acknowledge both the MIT Lincoln Laboratory Supercomputing Center (LLSC) and MIT SuperCloud teams for their help on tasks related to high performance computing. The authors also wish to thank Kevin Brady and Pooya Khorrami at MIT Lincoln Laboratory for their input on the active learning framework at the start of the project.

\bibliographystyle{IEEEtran}
\bibliography{IEEEabrv,main}

\begin{thebibliography}{10}
\providecommand{\url}[1]{#1}
\csname url@samestyle\endcsname
\providecommand{\newblock}{\relax}
\providecommand{\bibinfo}[2]{#2}
\providecommand{\BIBentrySTDinterwordspacing}{\spaceskip=0pt\relax}
\providecommand{\BIBentryALTinterwordstretchfactor}{4}
\providecommand{\BIBentryALTinterwordspacing}{\spaceskip=\fontdimen2\font plus
\BIBentryALTinterwordstretchfactor\fontdimen3\font minus
  \fontdimen4\font\relax}
\providecommand{\BIBforeignlanguage}[2]{{%
\expandafter\ifx\csname l@#1\endcsname\relax
\typeout{** WARNING: IEEEtran.bst: No hyphenation pattern has been}%
\typeout{** loaded for the language `#1'. Using the pattern for}%
\typeout{** the default language instead.}%
\else
\language=\csname l@#1\endcsname
\fi
#2}}
\providecommand{\BIBdecl}{\relax}
\BIBdecl

\bibitem{silvestri_confocal_2012}
L.~Silvestri, A.~Bria, L.~Sacconi, G.~Iannello, and F.~S. Pavone, ``Confocal
  light sheet microscopy: micron-scale neuroanatomy of the entire mouse
  brain,'' \emph{Optics Express}, vol.~20, no.~18, p. 20582, Aug. 2012.

\bibitem{chung2013clarity}
K.~Chung and K.~Deisseroth, ``Clarity for mapping the nervous system,''
  \emph{Nature methods}, vol.~10, no.~6, p. 508, 2013.

\bibitem{osten_mapping_2013}
P.~Osten and T.~W. Margrie, ``Mapping brain circuitry with a light
  microscope,'' \emph{Nature methods}, vol.~10, no.~6, p. 515, 2013, publisher:
  Nature Publishing Group.

\bibitem{tomer_advanced_2014}
R.~Tomer, L.~Ye, B.~Hsueh, and K.~Deisseroth, ``Advanced {CLARITY} for rapid
  and high-resolution imaging of intact tissues,'' \emph{Nature protocols},
  vol.~9, no.~7, p. 1682, 2014, publisher: Nature Publishing Group.

\bibitem{keller_visualizing_2015}
P.~J. Keller and M.~B. Ahrens, ``Visualizing {Whole}-{Brain} {Activity} and
  {Development} at the {Single}-{Cell} {Level} {Using} {Light}-{Sheet}
  {Microscopy},'' \emph{Neuron}, vol.~85, no.~3, pp. 462--483, Feb. 2015.

\bibitem{stefaniuk_light-sheet_2016}
M.~Stefaniuk, E.~J. Gualda, M.~Pawlowska, D.~Legutko, P.~Matryba, P.~Koza,
  W.~Konopka, D.~Owczarek, M.~Wawrzyniak, P.~Loza-Alvarez, and {others},
  ``Light-sheet microscopy imaging of a whole cleared rat brain with
  {Thy1}-{GFP} transgene,'' \emph{Scientific reports}, vol.~6, p. 28209, 2016,
  publisher: Nature Publishing Group.

\bibitem{fornito_fundamentals_2016}
A.~Fornito, A.~Zalesky, and E.~Bullmore, \emph{Fundamentals of brain network
  analysis}.\hskip 1em plus 0.5em minus 0.4em\relax Academic Press, 2016.

\bibitem{peng_extensible_2014}
H.~Peng, A.~Bria, Z.~Zhou, G.~Iannello, and F.~Long, ``Extensible visualization
  and analysis for multidimensional images using {Vaa3D},'' \emph{Nature
  protocols}, vol.~9, no.~1, p. 193, 2014, publisher: Nature Publishing Group.

\bibitem{peng_bigneuron_2015}
H.~Peng, M.~Hawrylycz, J.~Roskams, S.~Hill, N.~Spruston, E.~Meijering, and
  G.~A. Ascoli, ``{BigNeuron}: large-scale {3D} neuron reconstruction from
  optical microscopy images,'' \emph{Neuron}, vol.~87, no.~2, pp. 252--256,
  2015, publisher: Elsevier.

\bibitem{li20193d}
Q.~Li and L.~Shen, ``{3D} neuron reconstruction in tangled neuronal image with
  deep networks,'' \emph{IEEE Transactions on Medical Imaging}, 2019.

\bibitem{li2017deep}
R.~Li, T.~Zeng, H.~Peng, and S.~Ji, ``Deep learning segmentation of optical
  microscopy images improves {3-D} neuron reconstruction,'' \emph{IEEE
  Transactions on Medical Imaging}, vol.~36, no.~7, pp. 1533--1541, 2017.

\bibitem{wang2019multiscale}
H.~Wang, D.~Zhang, Y.~Song, S.~Liu, H.~Huang, M.~Chen, H.~Peng, and W.~Cai,
  ``Multiscale kernels for enhanced u-shaped network to improve 3d neuron
  tracing,'' in \emph{Proceedings of the IEEE Conference on Computer Vision and
  Pattern Recognition Workshops}, 2019.

\bibitem{he2017mask}
K.~He, G.~Gkioxari, P.~Doll{\'a}r, and R.~Girshick, ``Mask {R-CNN},'' in
  \emph{Proceedings of the IEEE international conference on computer vision},
  2017, pp. 2961--2969.

\bibitem{johnson2018adapting}
J.~W. Johnson, ``Adapting mask-rcnn for automatic nucleus segmentation,''
  \emph{arXiv preprint arXiv:1805.00500}, 2018.

\bibitem{ronneberger2015u}
O.~Ronneberger, P.~Fischer, and T.~Brox, ``U-net: Convolutional networks for
  biomedical image segmentation,'' in \emph{International Conference on Medical
  image computing and computer-assisted intervention}.\hskip 1em plus 0.5em
  minus 0.4em\relax Springer, 2015, pp. 234--241.

\bibitem{cciccek20163d}
{\"O}.~{\c{C}}i{\c{c}}ek, A.~Abdulkadir, S.~S. Lienkamp, T.~Brox, and
  O.~Ronneberger, ``{3D U-Net: Learning dense volumetric segmentation from
  sparse annotation},'' in \emph{International conference on medical image
  computing and computer-assisted intervention}.\hskip 1em plus 0.5em minus
  0.4em\relax Springer, 2016, pp. 424--432.

\bibitem{vuola2019mask}
A.~O. Vuola, S.~U. Akram, and J.~Kannala, ``Mask-{RCNN} and {U}-net ensembled
  for nuclei segmentation,'' \emph{arXiv preprint arXiv:1901.10170}, 2019.

\bibitem{ljosa2012annotated}
V.~Ljosa, K.~L. Sokolnicki, and A.~E. Carpenter, ``Annotated high-throughput
  microscopy image sets for validation.'' \emph{Nature methods}, vol.~9, no.~7,
  pp. 637--637, 2012.

\bibitem{januszewski2018high}
M.~Januszewski, J.~Kornfeld, P.~H. Li, A.~Pope, T.~Blakely, L.~Lindsey,
  J.~Maitin-Shepard, M.~Tyka, W.~Denk, and V.~Jain, ``High-precision automated
  reconstruction of neurons with flood-filling networks,'' \emph{Nature
  Methods}, vol.~15, no.~8, pp. 605--610, 2018.

\bibitem{khorrami2019deep}
P.~Khorrami, K.~Brady, M.~Hernandez, L.~Gjesteby, S.~N. Burke, D.~G. Lamb,
  M.~A. Melton, K.~J. Otto, and L.~J. Brattain, ``Deep learning-based nuclei
  segmentation of cleared brain tissue,'' in \emph{2019 IEEE High Performance
  Extreme Computing Conference (HPEC)}.\hskip 1em plus 0.5em minus 0.4em\relax
  IEEE, 2019, pp. 1--2.

\bibitem{settles_active_2009}
B.~Settles, ``Active learning literature survey,'' University of
  Wisconsin-Madison Department of Computer Sciences, Tech. Rep., 2009.

\bibitem{yang2017suggestive}
L.~Yang, Y.~Zhang, J.~Chen, S.~Zhang, and D.~Z. Chen, ``Suggestive annotation:
  A deep active learning framework for biomedical image segmentation,'' in
  \emph{International conference on medical image computing and
  computer-assisted intervention}.\hskip 1em plus 0.5em minus 0.4em\relax
  Springer, 2017, pp. 399--407.

\bibitem{kuo2018cost}
W.~Kuo, C.~H{\"a}ne, E.~Yuh, P.~Mukherjee, and J.~Malik, ``Cost-sensitive
  active learning for intracranial hemorrhage detection,'' in
  \emph{International Conference on Medical Image Computing and
  Computer-Assisted Intervention}.\hskip 1em plus 0.5em minus 0.4em\relax
  Springer, 2018, pp. 715--723.

\bibitem{schneider2012nih}
C.~A. Schneider, W.~S. Rasband, and K.~W. Eliceiri, ``{NIH Image to ImageJ}: 25
  years of image analysis,'' \emph{Nature methods}, vol.~9, no.~7, pp.
  671--675, 2012.

\bibitem{morgan1987mapping}
J.~I. Morgan, D.~R. Cohen, J.~L. Hempstead, and T.~Curran, ``Mapping patterns
  of c-fos expression in the central nervous system after seizure,''
  \emph{Science}, vol. 237, no. 4811, pp. 192--197, 1987.

\bibitem{prout2017supercloud}
A.~Prout, W.~Arcand, D.~Bestor, B.~Bergeron, C.~Byun, V.~Gadepally, M.~Hubbell,
  M.~Houle, M.~Jones, P.~Michaleas \emph{et~al.}, ``Mit supercloud portal
  workspace: Enabling hpc web application deployment,'' in \emph{2017 IEEE High
  Performance Extreme Computing Conference (HPEC)}.\hskip 1em plus 0.5em minus
  0.4em\relax IEEE, 2017, pp. 1--6.

\bibitem{yoo2003slurm}
A.~B. Yoo, M.~A. Jette, and M.~Grondona, ``Slurm: Simple linux utility for
  resource management,'' in \emph{Workshop on Job Scheduling Strategies for
  Parallel Processing}.\hskip 1em plus 0.5em minus 0.4em\relax Springer, 2003,
  pp. 44--60.

\bibitem{travinin2007pmatlab}
N.~Travinin~Bliss and J.~Kepner, ``{pMATLAB Parallel MATLAB Library},''
  \emph{The International Journal of High Performance Computing Applications},
  vol.~21, no.~3, pp. 336--359, 2007.

\bibitem{meyer1994topographic}
F.~Meyer, ``Topographic distance and watershed lines,'' \emph{Signal
  processing}, vol.~38, no.~1, pp. 113--125, 1994.

\end{thebibliography}

\end{document}